\documentclass{PoS}

\title{Origin of a large CP asymmetry in \mbox{{\boldmath $B^{\pm} \rightarrow K^+ K^- K^{\pm}$}}
 decays}
\ShortTitle{CP asymmetry in $B^{\pm} \rightarrow K^+ K^- K^{\pm}$ }

\author{\speaker{Leonard Le\'sniak} and Piotr \.Zenczykowski \\
        Division of Theoretical Physics, The Henryk Niewodnicza\'nski Institute of Nuclear Physics,\\ 
Polish Academy of Sciences, 31-342, Krak\'ow, Poland\\
        E-mail: \email{Leonard.Lesniak@ifj.edu.pl}}

\abstract{Large $CP$-violating asymmetry effects in the 
$B^{\pm} \rightarrow K^+ K^- K^{\pm}$ decays 
have been predicted in the QCD factorization model.
The model includes strong $K^+ K^-$ final-state long-distance interactions in
the $S$-, $P$-, and also (in the recent analysis) $D$- wave two-body states.
The $S$-wave two-body unitarity conditions involve interchannel couplings of
the kaon-kaon states with the intermediate states of two pions and four pions.
As a result, the pion-pion to kaon-kaon rescattering effects are included in the
model. 
It is shown how the weak phase differences together with the existence of
two different strong phases of the $S$-wave decay amplitudes (related to the phases of the
kaon scalar strange and non-strange form factors) contribute to the $CP$-asymmetry in question.
The theoretical results are compared with recent experimental 
data of the LHCb and BABAR Collaborations.
    }

\FullConference{The European Physical Society Conference on High Energy Physics\\
		22--29 July 2015\\
		Vienna, Austria}

\begin{document}

\section{Decay amplitudes}
Large $CP$ asymmetry in $B^{\pm} \rightarrow K^+ K^- K^{\pm}$ decays for the 
$K^+ K^-$ effective masses above the $\phi(1020)$ meson peak has been predicted in a model 
developed in 2011 \cite{FKLZ}. In 2013 the LHCb Collaboration measured the $CP$-violating 
asymmetry and found a large negative asymmetry in the region 
$1.1 < m^2_{K^+ K^–~ low} < 2.0$ GeV$^2$ and
 $m^2_{K^+ K^–~  high} < 15$ GeV$^2$ \cite{LHCb}.
A similar result has been also obtained 
in an earlier BABAR measurement \cite{BABAR}. The above data from two
collaborations have been recently analysed in Ref.~\cite{LZ}. Here we explain in more detail
the origin of this large negative $CP$ asymmetry.

 The $CP$-violating asymmetry, measured in \cite{LHCb,BABAR}, is small in the range of the $K^+ K^-$
effective masses under
the $\phi(1020)$ meson maximum where the $P$-wave $K^+ K^-$ final state strong 
interaction apparently dominates. Thus we analyse properties of the
amplitudes $A^-_S$ and  $A^+_S$ corresponding respectively to the 
$B^-\rightarrow K^+ K^- K^-$ and $B^+\rightarrow K^+ K^- K^+$ decays in which the 
$K^+ K^-$ pair is formed in the $S$-wave. The decay amplitudes are calculated in the quasi-two-body
QCD factorization model \cite{FKLZ}. There are two weak decay transition amplitudes proportional 
to the following two products of the elements of the Cabibbo-Kobayashi-Maskawa (CKM) quark mixing 
matrix:
$\lambda_u=V_{ub}V^*_{us}$ and $\lambda_c=V_{cb}V^*_{cs}$. The $B^-\rightarrow K^+ K^- K^-$ 
decay amplitude is equal to
\begin{equation}
 A^-_S=\lambda_u F_u +\lambda_c F_c.
\label{A-}
\end{equation}
while the charge conjugated $B^+\rightarrow K^+ K^- K^+$ decay amplitude reads
\begin{equation}
 A^+_S=\lambda_u^* F_u+\lambda_c^* F_c.
\label{A+}
\end{equation}
In Eqs.~(\ref{A-}) and ~(\ref{A+}) the functions $F_u$ and $F_c$ include the weak
amplitude factors and the strong interaction factors. The strong factors do not change under 
the charge 
conjugation when one passes from the $B^-$ to the $B^+$ decay. With a good accuracy, the
$V^*_{us}, V_{cb}$ and $V^*_{cs}$ matrix elements are real, while $V_{ub}=|V_{ub}|e^{-i \gamma}$
is complex. The CKM angle $\gamma$ is close to 70 degrees. The modulus of the $\lambda_u$
coefficient is smaller than that of $\lambda_c$ (by a factor of about 51). The phase $\phi_c$
of  $\lambda_c$ is very close to zero while the phase $\phi_u$ of $\lambda_u$ is equal to 
$(-\gamma)$.
The full expressions for the $S$-wave decay amplitudes are given in \cite{LZ}. 
For a further discussion, it is useful to write the factors 
$F_u$ and $F_c$ in terms of the scalar non-strange kaon form factor $\Gamma_2^n$ and the
scalar strange kaon form factor $\Gamma_2^s$:
\begin{equation}
  F_u=u_n \Gamma_2^{n*} +u_s\Gamma_2^{s*}   
\label{Fu}
\end{equation} 
and 
\begin{equation}
  F_c=c_n \Gamma_2^{n*} +c_s\Gamma_2^{s*}.   
\label{Fc}
\end{equation} 
Here $u_n,u_s,c_n$ and $c_s$ are simple functions which can be read from Eq. (2) of 
~\cite{LZ}. 
In $F_u$ the first term dominates: $|u_n\Gamma_2^{n*}|\gg|u_s\Gamma_2^{s*}|$
as in $u_n$ the tree part of the weak amplitude is much larger than the penguin components
present in $u_s$.
In $F_c$ only the penguin terms contribute and one can verify that the coefficients $c_n$ and 
$c_s$ are comparable.

\section{Kaon scalar form factors}

   The final-state strong $K^+K^-$ interactions play an important role in the
description of the $CP$ asymmetry. In the present application we use the coupled channel 
model of the scalar-isoscalar form factors. Three coupled channels are considered:
 $\pi\pi$ ($\pi^+\pi^-$ and $\pi^0\pi^0$) - labelled below by subscript 1, $K \bar K$ 
($K^+K^-$ and $K^0 \bar K^0$ - subscript 2
and the effective 4$\pi$ channel (quasi-two body $\sigma\sigma$ or $\rho\rho$) - subscript 3. 
Then the six scalar form factors are defined as follows:
\begin{equation}
  \Gamma^{n}= \left( \begin{array}{c}
              \Gamma^{n}_1 \\
              \Gamma^{n}_2 \\
              \Gamma^{n}_3
                        \end{array} \right), ~~~~~~~~ \Gamma^{s}= \left( \begin{array}{c}
              \Gamma^{s}_1 \\
              \Gamma^{s}_2 \\
              \Gamma^{s}_3
                        \end{array} \right).  
\label{Gamma}
\end{equation} 
The superscript $n$ labels the non-strange form factors, while $s$ - the strange form
factors. Below we list nine meson-meson scattering amplitudes $T$ which should be 
parametrized before the meson form factors are calculated: 
\begin{equation}
T= \left( \begin{array}{ccc} 
           T_{11} & T_{12} & T_{13} \\
           T_{21} & T_{22} & T_{23} \\
           T_{31} & T_{32} & T_{33} 
          \end{array} \right).  
\label{T}
\end{equation}
The diagonal elements of $T$ are elastic scattering amplitudes, for example $T_{22}$ is the elastic
$S$-wave isoscalar $KK$ amplitude. The non-diagonal $T$ elements are the interchannel transition  
amplitudes.
The meson-meson amplitudes are calculated using a model described in \cite{KLL}.
In ~\cite{JPD} one can find a general derivation of the scalar pion form factors used in the 
studies of the $B^{\pm} \rightarrow \pi^+ \pi^- \pi^{\pm}$ decays.
The kaon form factors are discussed in \cite{FKLZ,LZ}.
Below we write a general matrix structure which shows the functional dependence of the scalar 
non-strange 
and strange form factors on the meson-meson scattering amplitudes:
\begin{equation}
\Gamma^{n*}=R^n+T G R^n,~~~~~~~~\Gamma^{s*}=R^s+T G R^s.
\label{gammans}
\end{equation}
Here $G$ is the diagonal matrix of the Green functions
\begin{equation}
G= \left( \begin{array}{ccc} 
           G_1  & 0   & 0 \\
           0    & G_2 & 0 \\
           0    & 0   & G_3 
          \end{array} \right).  
\label{Green}
\end{equation}
The $R^n$ and $R^s$ are matrices of the production functions whose three components are 
parametrized in the following way:
\begin{equation}
R^{n,s}_j(E)=\frac{\alpha^{n,s}_j+\tau^{n,s}_j E+ \omega^{n,s}_j E^2}{1+cE^4}, ~~~j=1,2,3,
\label{R}
\end{equation}
where the variable $E$ equals to the effective masses of two mesons in each channel:
\begin{equation}
E=m_{\pi\pi}=m_{KK}=m_{4\pi}.
\label{E}
\end{equation}
The real parameters $\alpha^{n,s}_j$, $\tau^{n,s}_j$ and $\omega^{n,s}_j$ are constrained using 
the chiral perturbation model and the fitted parameter $c$ controls the high energy behaviour
 of $R_j$.
In the integral equations ~(\ref{gammans}) one sums over all the appropriate coupled components
of the scattering matrix $T$. More specifically, the non-strange and strange kaon form factors 
are expressed as
\begin{equation}
\Gamma_2^{n*}=R_2^{n}+T_{22} G_2 R_2^{n} + T_{21} G_1 R_1^{n}+T_{23} G_3 R_3^{n}
\label{G2n}
\end{equation}
and
\begin{equation}
\Gamma_2^{s*}=R_2^{s}+T_{22} G_2 R_2^{s} + T_{21} G_1 R_1^{s} + T_{23} G_3 R_3^{s}.
\label{G2s}
\end{equation}
In these two equations the following three meson-meson $S$-wave amplitudes are present:
the $KK \to KK$ elastic scattering amplitude $T_{22}$, the $\pi \pi \to KK$ transition
(or rescattering) amplitude $T_{21}$ and the $4\pi \to KK$ transition amplitude $T_{23}$.
Since the non-strange production functions $R^n_j$ are different from the strange production
 functions $R^s_j$ and the rescattering amplitudes $T_{21}$ and $T_{23}$ are in general
non-zero, the complex functions $\Gamma_2^{s*}$ and $\Gamma_2^{n*}$ have different phases
($\delta_s \neq \delta_n$).

\begin{figure} 
\begin{center}

\includegraphics[scale = 0.27,angle=270]{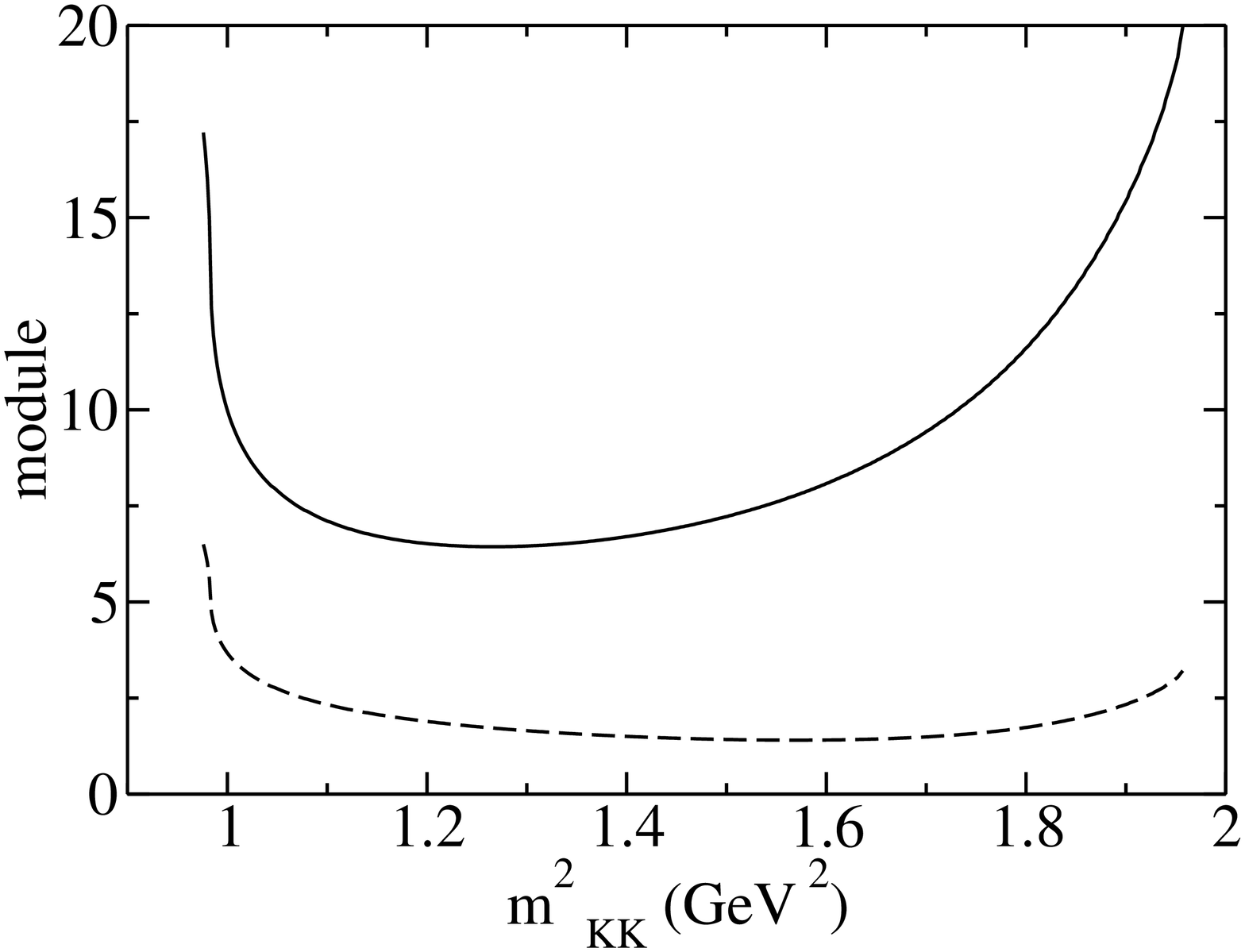}~
\includegraphics[scale = 0.27,angle=270]{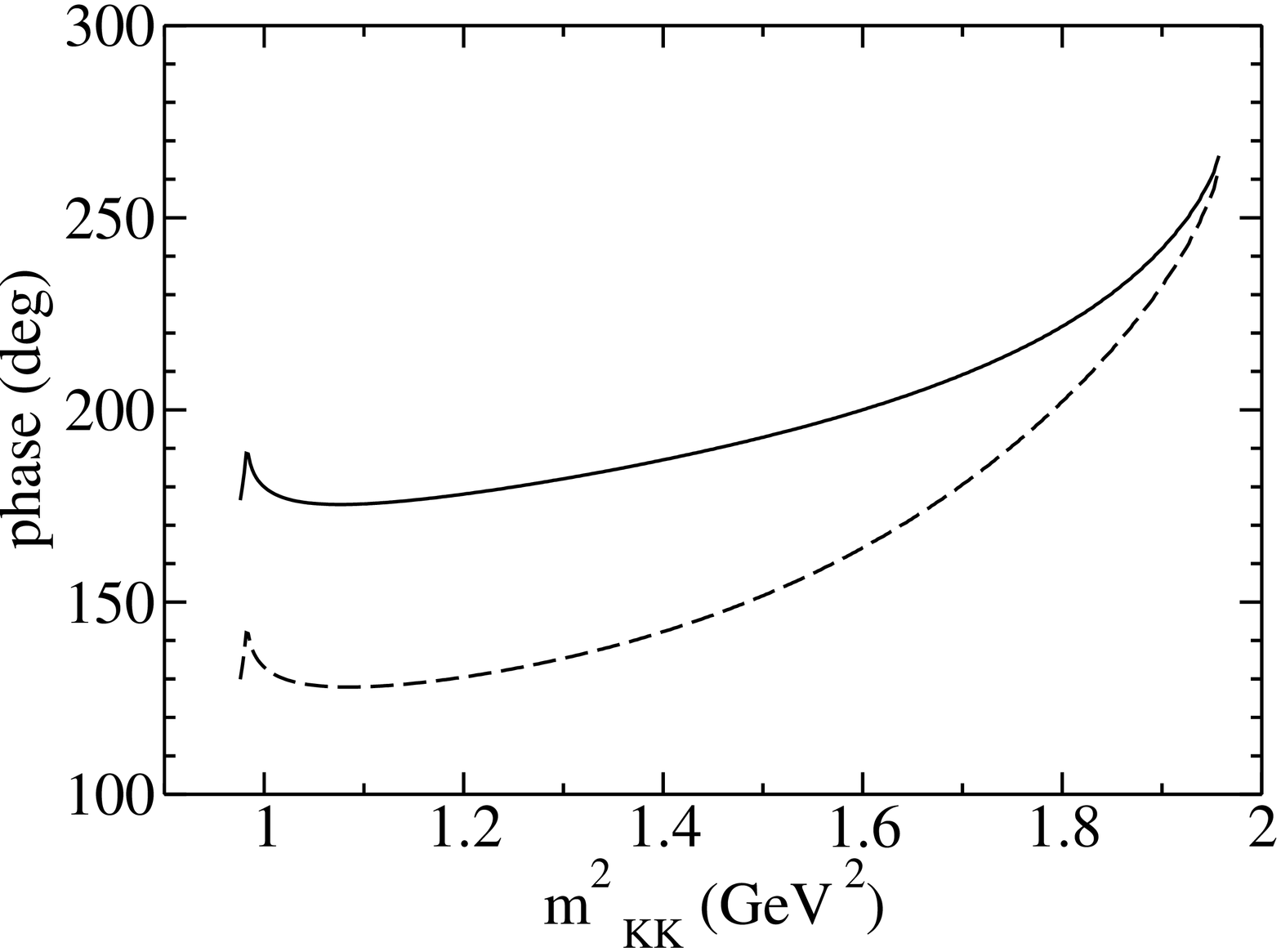}~~~~

\caption{The moduli (left panel) and the phases (right panel) of the 
kaon non-strange and strange scalar form factors $\Gamma^{n*}_2$ and $\Gamma^{s*}_2$
(solid and dashed lines, respectively).
}
\label{Formf}
\end{center}
\end{figure}

The moduli and the phases of the kaon form factors $\Gamma_2^{n*}$ and $\Gamma_2^{s*}$
are shown in Fig. \ref{Formf}.

 Near the effective $K^+K^-$ mass close to 1 GeV one sees
maxima of the moduli and the phases. 
This behaviour is related to the presence of the $f_0(980)$ resonance. 
One should also notice that the modulus of the non-strange form factor is a few 
times larger that the modulus of the strange form factor. 
This inequality contributes to the smallness of the ratio $|F_c/F_u|$.
In a wide range of the effective masses above the position
 of the $f_0(980)$ resonance
the phases $\delta_n$ and $\delta_s$ differ by about 45 degrees,
 leading in consequence
to a large negative asymmetry.
    
\section{$CP$ asymmetry}

In a region of the effective $K^+K^-$ masses, where the $S$-wave part of the decay amplitude
dominates, the $CP$ asymmetry can be written as
\begin{equation}
  A_{CP}=\frac{|A^-_S|^2 -  |A^+_S|^2}{|A^-_S|^2 +  |A^+_S|^2} = 
-\frac{2rsin(\phi_c-\phi_u) sin(\delta_c-\delta_u)}{1+r^2+2rcos(\phi_c-\phi_u) 
cos (\delta_c-\delta_u)}, 
\label{ACP}
\end{equation} 
where $\phi_u$ and $\phi_c$ are the phases of functions $F_u$ and $F_c$, respectively and the 
factor $r$ is defined as the ratio $|\frac{\lambda_c F_c}{\lambda_u F_u}|$.
Let us define phases $\delta_n$ and $\delta_s$ of the non-strange and strange kaon form factors:
$\Gamma_2^{n*}=|\Gamma_2^{n}|e^{i\delta_n}$ and $\Gamma_2^{s*}=|\Gamma_2^{s}|e^{i\delta_s}$.
Then from the functional dependence of the factors $F_u$ and $F_c$ we can derive an approximate
relation
\begin{equation}
  \delta_c - \delta_u \approx \delta_s- \delta_n.
\label{diff}
\end{equation} 
One can also calculate numerically the ratio $|F_c|/|F_u| \approx 0.02$ for 
$ m^2_{K^+ K^–~ low} \approx 1.2$ GeV$^2$. 
Recalling that $\lambda_c/\lambda_u \approx 51$, one obtains
 $r \approx 1$. 
As the inspection of the form of Eq. (3.1) shows, $r=1$ is 
the value for which 
the modulus of the $CP$ asymmetry reaches its maximal value. 
Since $\phi_c-\phi_u=\gamma \approx 68^0$
and for $1.0 <  m^2_{K^+ K^–~ low} <1.5$ GeV$^2$ the difference $\delta_s-\delta_n$ 
is approximately $45^0$, in this range of
effective $K^+K^-$ masses the $CP$ asymmetry reaches a large negative
 value of about $-0.5$. 

In addition to the $S$-wave part of the decay amplitude
discussed above, the $P$- and $D$- waves of the $K^+K^-$ scattering are also included in the  
model.
The full model has five parameters fitted from the data.

\begin{figure} 
\begin{center}

\includegraphics[scale = 0.4]{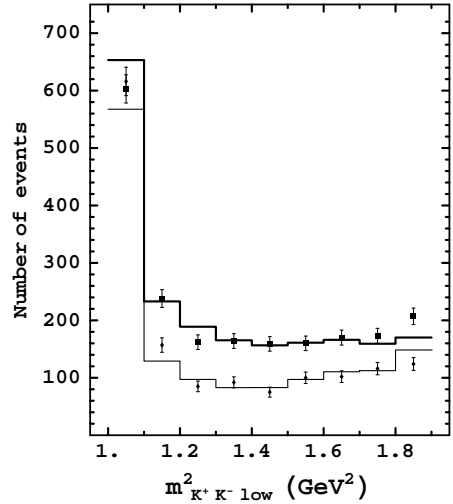}
\caption{Distribution of the LHCb signal events for the $B^{\pm} \rightarrow K^+ K^- K^{\pm}$ 
decays. The data points are from \cite{LHCb} ($B^+$ - squares, $B^-$ - diamonds). The model results
are shown as thick ($B^+$) and thin ($B^-$) histograms.}
\label{daneN}
\end{center}
\end{figure}

\begin{figure} 
\begin{center}

\includegraphics[scale = 0.6]{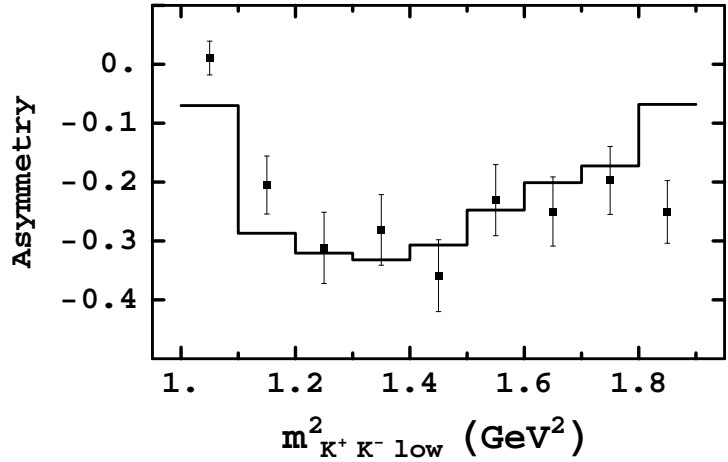}

\caption{Distribution of the $CP$- violating 
asymmetry for the $B^{\pm} \rightarrow K^+ K^- K^{\pm}$ decays.
The data points are from \cite{LHCb}, the model results are shown as solid histogram in bins 
of 0.1 GeV$^2$.} 
\label{daneACP}
\end{center}
\end{figure}

The results of the fits are compared with the LHCb data in Figs.~\ref{daneN} and \ref{daneACP}.
In the data presented in 
Fig.~\ref{daneN}, except for the first bin dominated by the $\phi(1020)$ meson contribution, 
one observes a significant
surplus of the number of events corresponding to the $B^+$ decays over the number of events coming
 from the 
$B^-$ decays. This is also visualized in Fig.~\ref{daneACP} where the negative $CP$ asymmetry 
is shown up to $m^2_{K^+ K^–~ low}$ equal to about 1.9 GeV$^2$. The $CP$ asymmetry data of the BABAR 
Collaboration \cite{BABAR} can also be fitted using the theoretical model of Ref.~\cite{LZ}.

\section{Short summary}

Large CP-violating asymmetry effects in the $B^{\pm} \rightarrow K^+ K^- K^{\pm}$
 decays have been predicted in the QCD factorization model.
The model includes strong $K^+ K^-$ final-state long-distance interactions in the 
$S$-, $P$- and $D$- waves.
We have shown how a large negative $CP$ asymmetry appears for the $K^+ K^-$ effective
 mass squares between 1.1 and 1.9 GeV$^2$ . 
The asymmetry originates mainly in the $S$-wave amplitudes. It stems from the presence of two
significantly differing weak phases together with the existence of two 
different strong phases related 
to the phases of the kaon scalar strange and non-strange form factors.
 With the $\pi\pi \rightarrow KK$ and $4\pi \rightarrow KK$ rescattering effects included in the model,
 the interchannel couplings of the kaon-kaon state to the two pion and four pion states 
naturally produce the difference in the relevant strong phases.


\section{Acknowledgements}
This work has been partially supported by the Polish National Science Centre\\
(grant no 2013/11/B/ST2/04245).

\end{document}